\documentstyle[epsf]{cupconf}

\ifoldfss
\else
  \ifnfssone
    \newmathalphabet{\mathit}
      \addtoversion{normal}{\mathit}{cmr}{m}{it}
      \addtoversion{bold}{\mathit}{cmr}{bx}{it}
    \newmathalphabet{\mathcal}
      \addtoversion{normal}{\mathcal}{cmsy}{m}{n}
    \else
    \ifnfsstwo
    \fi
  \fi
\fi

\def\hexnumber#1{\ifcase#1 0\or1\or2\or3\or4\or5\or6\or7\or8\or9\or
 A\or B\or C\or D\or E\or F\fi }

\makeatletter
\ifx\CUP@mtlplain@loaded\undefined
\else
\fi
\makeatother
 \makeatletter
 \ifx\CUP@mtlplain@loaded\undefined
   \font\tenbmi=cmmib10 at 10pt
   \font\sevenbmi=cmmib10 at 7pt
   \font\fivebmi=cmmib10 at 5pt

   \newfam\bmifam
   \textfont\bmifam=\tenbmi
   \scriptfont\bmifam=\sevenbmi
   \scriptscriptfont\bmifam=\fivebmi
   
 \fi
 \makeatother
\ifnfsstwo

\fi
\ifnfssone

\fi
\ifoldfss

\fi

\mathchardef\varLambda="0103

\makeatletter
\ifx\CUP@mtlplain@loaded\undefined
\else
\fi
\makeatother
\makeatletter
\ifx\CUP@mtlplain@loaded\undefined
  \font\tenbms=cmbsy10
  \font\sevenbms=cmbsy10 at 7pt
  \font\fivebms=cmbsy10 at 5pt
  \newfam\bmsfam
  \textfont\bmsfam=\tenbms
  \scriptfont\bmsfam=\sevenbms
  \scriptscriptfont\bmsfam=\fivebms

  \edef\bsy@{\hexnumber\bmsfam}
  \mathchardef\bnabla="0\bsy@72
\fi
\makeatother
\title[Focussing in on H$_0$]{Focussing in on H$_0$\\
\normalsize \bf (Dedicated to Allan Sandage for his seventieth birthday)}

\author[G.A~Tammann \& M.~Federspiel]%
{G.\ns A.\ns T\ls A\ls M\ls M\ls A\ls N\ls N\ls$^1$,\ns 
\and \ns M.\ns F\ls E\ls D\ls E\ls R\ls S\ls P\ls I\ls E\ls L\ls$^1$}

\affiliation{$^1$Astronomisches Institut der Universit\"at Basel,
Venusstr.~7, CH-4102 Binningen, Switzerland}

\begin{document}
\ifnfssone
\else
  \ifnfsstwo
  \else
    \ifoldfss
      \let\mathcal\cal
      \let\mathrm\rm
      \let\mathsf\sf
    \fi
  \fi
\fi

\maketitle

\begin{abstract}
Six independent methods yield a Virgo cluster modulus of $(m-M)=31.66\pm 
0.08$. This inserted into the Hubble diagram of clusters, whose relative
distances to the Virgo cluster are known, gives $H_0=54\pm 4$. The result
is independent of any observed or inferred velocity of the Virgo cluster
and holds out to $\sim 10\,000\,$km s$^{-1}$. An analogous use of the Fornax
cluster is not yet possible because of its unknown structure in space
and its poorly understood distance relative to other clusters. Field
galaxies, corrected for Malmquist bias, give $H_0=53\pm 3$. The results
of purely physical distance determinations cluster around $H_0=55\pm 10$.
These results may be compared with the high-weight determination of
$H_0=58.5\pm 3$ from Cepheid-calibrated SNe$\,$Ia and their Hubble diagram
out to $\sim 30\,000\,$km s$^{-1}$ (Saha 1996). The agreement of the 
independent methods suggests that the external error of an adopted value
of $H_0=58$ is equal or less than 13\%.
\end{abstract}
\firstsection % if your document starts with a section,
\section{Introduction}
The determination of distances of individual galaxies and the calibration
of the Hubble constant $H_0$ are two connected, but fundamentally different
problems. The calibration of $H_0$ requires in addition to reliable galaxy
distances the demonstration that the galaxies under consideration do partake
of the expansion of the Universe and that their velocities are not
dominated by, for instance, peculiar or virial motions. Moreover, the
question must be answered whether the expansion can be characterized by
a single value of $H_0$, or whether $H_0$ is scale-variant or even a
stochastic variable. 

The ensuing strategy is clear. One must construct a Hubble diagram plotting
for suitable objects (e.g.~SNe$\,$Ia, galaxies, or clusters) $\log cz$
versus $\log$ ({\sl relative} distance). The relative distance is
approximated by the apparent magnitude of \lq\lq standard candles\rq\rq .
(Standard candles are to be objects whose scatter in absolute magnitude
is $<0.3\,$mag to avoid gross effects of Malmquist bias.) Alternatively,
relative distances of non-standard candles, for instance clusters, can
come from various distance indicators, whose zero-point calibration is
not needed. In either case the resulting Hubble diagram provides two
basic informations. (1) Do the objects under consideration partake of
the {\sl linear} expansion (corresponding to a ridge line  slope of 0.2!),
and (2) what is the magnitude scatter about the ridge line? This scatter
is decisive for the precision with which the intercept of the ridge line
-- and hence $H_0$ -- can be determined.

Once the Hubble diagram is found satisfactory out to a maximum recession
velocity, the next step is to determine the absolute magnitude or the
linear distance of one or more objects defining this Hubble diagram.
The resulting value of $H_0$ is valid then out to the same maximum
recession velocity. 

The most successful application of this strategy to date comes from the Hubble
diagram of SNe$\,$Ia, which is defined out to $30\,000$ km s$^{-1}$ -- and
hence provides a truly cosmic value of $H_0$ -- and its calibration by
means of seven SNe$\,$Ia with known Cepheid distances (Saha 1996).

An independent application of essentially the same strategy is discussed
in Section 2. Here a Hubble diagram is constructed from 31 clusters whose
distances are known {\sl relative} to the Virgo cluster. By inserting the
best available distance of the latter one can transform the Hubble diagram
into a diagram of recession velocity versus {\sl linear} distance, allowing
to read off $H_0$ at any distance up to $11\,000$ km s$^{-1}$.

In Section 3 the still poorly determined distance of the Fornax cluster
as well as the evidence from the Coma cluster are discussed. The evidence of 
$H_0$ from field galaxies and physical distance determinations is compiled 
in Sections 4 and 5. The conclusions follow in Section 6.
\section{The Global Value of H$_0$ from the Virgo Cluster Distance}
\subsection{The Hubble Diagram of Clusters out to 11$\,$000 km s$^{-1}$}

Cluster distances relative to the Virgo cluster are available for 17
clusters from various methods like the Tully-Fisher (TF) method, the
$D_{\rm n}-\sigma$ relation, and first-ranked cluster galaxies (for a
compilation see Jerjen \& Tammann 1993). In addition, carefully
determined relative TF distances are available for 24 clusters from
\cite{Gio96a}. The latter list does not include the Virgo cluster, but
since eight clusters are in common, the two lists can be combined with
a mean error of only 0.05 mag. The resulting list of 31 cluster
distances relative to the Virgo cluster is given in
Table~\ref{tab:clusters}. The double cluster A$\,$2634/66 is not used
here because Giovanelli only gives a distance for A$\,$2634.

\begin{table}
\caption{Mean cluster distances relative to the Virgo cluster}
\label{tab:clusters}
\begin{center} \footnotesize
\begin{tabular}{lrrcl}
\hline
%\noalign{\smallskip}
 cluster & $v_{220}$ & $v^{\rm CMB}$ & $(m-M)-(m-M)_{\rm Virgo}$ & 
 source$^{\ast}$ \\
%\noalign{\smallskip}
\hline
%\noalign{\smallskip}
 Ursa Major & 1270 &  & 0.37 & JT, G \\
 Fornax & 1338 &  & 0.23 & JT, G \\
 Eridanus & 1522 & & 0.93 & G \\
 MDL$\,$59 & 2636 & & 1.85 & G \\
 Antlia & 2767 & & 1.93 & G \\
 N$\,$3557 & 2973 & & 2.08 & G \\
 ESO$\,$508 & 3029 & & 1.83 & G \\
 Centaurus & 3043 &  & 2.18 & JT, G \\
 Pegasus &  & 3517 & 2.46 & JT, G\\
 Hydra I &  & 4050 & 2.73 & JT, G \\
 Pavo & & 4055 & 2.60 & G \\
 Pavo II & & 4444 & 2.78 & G \\
 A$\,$262 & & 4664 & 2.92 & G \\
 Pisces &  & 4717 & 2.99 & JT \\
 N$\,$507 & & 4808 & 2.97 & G \\
 A$\,$3574 & & 4817 & 3.03 & G \\
 N$\,$383 & & 4865 & 3.04 & G \\
 Cancer &  & 5026 & 3.07 & JT, G\\
 Perseus &  & 5178 & 3.23 & JT \\
 Zw74-23 &  & 6308 & 3.60 & JT \\
 A$\,$1367 &  & 6777 & 3.72 & JT, G\\
 A$\,$400 &  & 6920 & 3.95 & JT, G\\
 A$\,$1656 & & 7185 & 3.84 & G \\
 Coma &  & 7188 & 3.80 & JT \\
 A$\,$539 &  & 8630 & 4.39 & JT\\
 A$\,$2634/66 &  & 8265 & 4.16 & JT, G \\
 A$\,$2199 & & 8996 & 4.45 & G \\
 A$\,$2197 & & 9162 & 4.45 & G \\ 
 A$\,$1185 &  & 10808 & 5.03 & JT\\
 A$\,$2147 &  & 10981 & 4.98 & JT \\
 Hercules &  & 11058 & 4.71 & JT \\
\multicolumn{5}{l}{$^{\ast}$sources: JT$ = $\cite{JT93}; G$ = $Giovanelli 
 (1996a)} \\
\end{tabular}
\end{center}
\end{table}

The Hubble diagram of the 31 clusters is shown in Fig.~1. Clusters with
$v_0 < 3000\,$ km s$^{-1}$ are corrected for a Virgocentric infall model
with a local infall velocity of $220\,$ km s$^{-1}$. More distant clusters
do not partake of the local motion with respect to the CMB. They are
therefore corrected for a CMB vector of $630\,$ km s$^{-1}$. The above
dividing limit of $3000\,$ km s$^{-1}$ is an educated guess (cf.~Jerjen \&
Tammann 1993, Giovanelli 1996a); the exact choice has no effect on the
following conclusions.

\begin{figure} 
%\vspace{10cm}
\begin{center}
\leavevmode
\epsfxsize 12.8cm
\epsffile{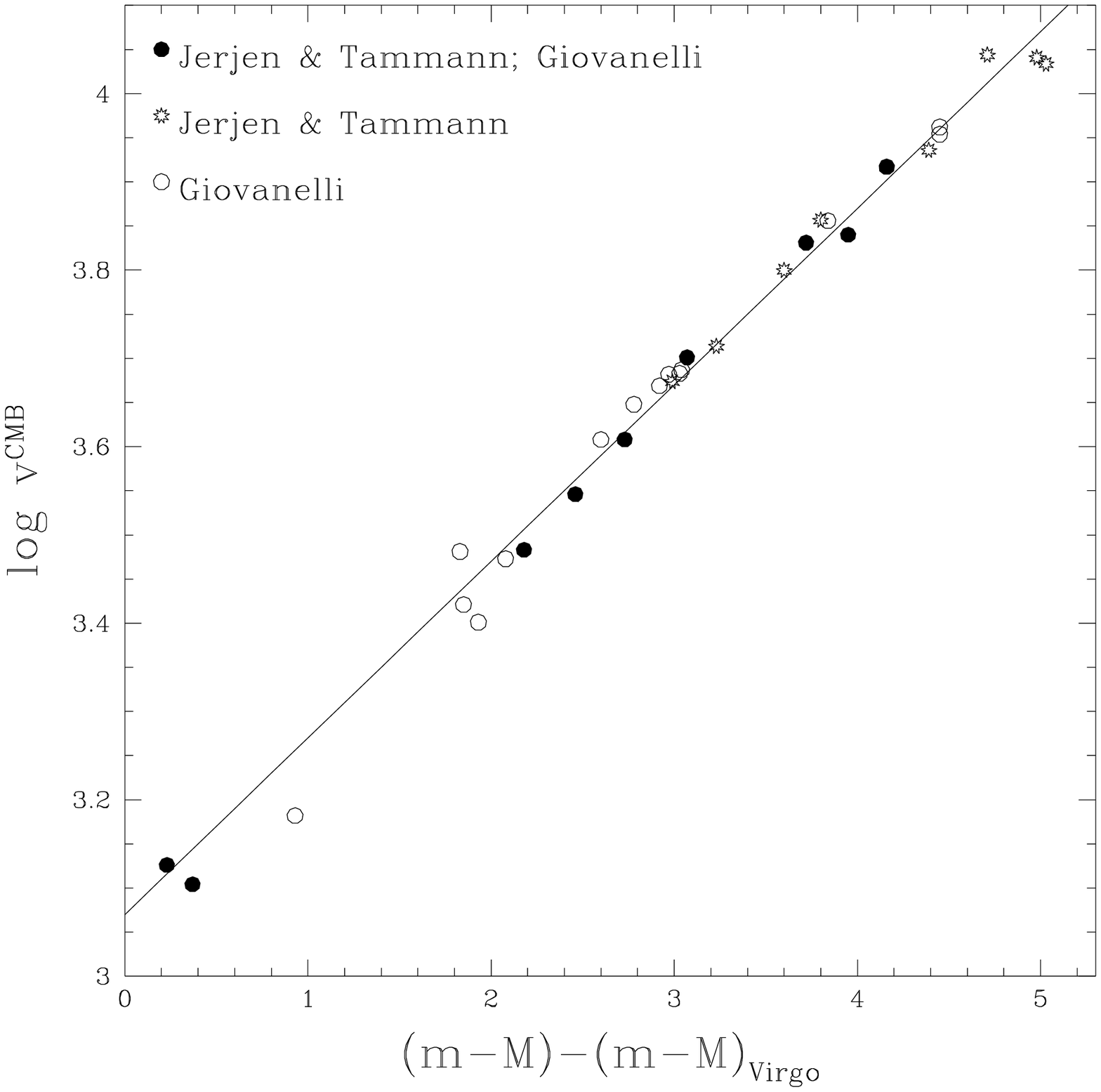}
\caption{Hubble diagram of 31 clusters with known relative
distances. The data were taken from Jerjen \& Tammann (1993;
asterisks) and Giovanelli (1996a, open circles). Nine clusters are
listed in both sources (filled circles). The abscissa gives the
distance modulus relative to the Virgo cluster and the ordinate the
log of the recession velocity referred to the CMB. For \lq\lq
local\rq\rq ~clusters with $v_0<3000\,$km s$^{-1}$ the velocities are
referred to the centroid of the Local Group and corrected for
Virgocentric infall.} 
\end{center}
\end{figure} 

The ridge line in Fig.~1 is represented by
\begin{equation}
\log v^{\rm CMB}=0.2 \Bigl [(m-M)-(m-M)_{\rm Virgo}\Bigr ]+3.068\pm 0.024, 
\ \ \ \sigma=0.13.
\end{equation}

The slope of 0.2 is forced. A free fit gives a slope of $0.205\pm 0.005$ (or
$4.82\pm 0.20$ for the inverse regression) in statistically perfect agreement
with a {\sl linear} expansion.

Simple transformation of equation (1) gives
\begin{equation}
\log H_0=\log v^{\rm CMB}-\log r_{\rm Mpc}=-0.2 (m-M)_{\rm Virgo} + (8.068\pm 
0.024)
\end{equation}
Inserting the distance modulus of the Virgo cluster into equation (2.2) 
immediately yields the value of $H_0$ out to $11\,000\,$km s$^{-1}$.
The next section is therefore devoted to a discussion of the Virgo cluster
distance.

If the strongly deviating Eridanus cluster (cf.~Fig.~1) is excluded the
constant term in eq.~(2.1) changes only marginally to be $3.070\pm 0.020$
and the scatter reduces to $\sigma=0.11$ mag. This scatter restricts the
peculiar motions of the cluster centers quite severely. Even an optimistic
error of the relative distances of 0.08 mag leaves a scatter of only 0.07
mag due to peculiar motions, which translates into a mean (one-dimensional)
peculiar motion of $\sim 350\,$km s$^{-1}$ (cf.~Jerjen \& Tammann 1993).

It is noted in passing that eq.~(2.1) also predicts the Virgo cluster
velocity in the CMB frame at a relative distance of 0, i.e.~$1169\pm
30\,$km s$^{-1}$. This is the velocity one would observe in the
absence of all local peculiar or streaming velocities. Comparing this
value with the actually observed cluster velocity of $v_0=1050\pm
35\,$km s$^{-1}$ (Binggeli, Popescu \& Tammann 1993) or correspondingly $v_{\rm
LG}=922\pm 35\,$km s$^{-1}$, one obtains a Virgocentric infall
velocity of the Local Group of $v_{\rm infall}=247\pm 46\,$km s$^{-1}$
(cf.~Jerjen \& Tammann 1993). We take this as a confirmation of
$v_{\rm infall}=220\,$km s$^{-1}$ adopted above (Tammann and Sandage
1985). The result is also compatible with $v_{\rm infall}=275\pm
90\,$km s$^{-1}$ from the Hubble diagram of SNe$\,$Ia (Hamuy et
al.~1996). 

{\sl It must be stressed that the determination of $H_0$ from
eq.~(2.2) depends on the quality of the Virgo cluster distance, but
it is totally independent of any observed or inferred velocity of that
cluster.}

\subsection{The Virgo Cluster Distance from Cepheids}

Observations of Cepheids in Virgo cluster galaxies require the Hubble
Space Telescope (HST) because due to resolution problems essentially all
stars selected from the ground as \lq\lq single{\rq\rq } turn out to be
blends or multiples on HST images. 

The first HST Cepheid distance of a galaxy in the Virgo complex was for
NGC$\,$4321 (M$\,$100; Freedman et al.~1994; see also Ferrarese et al.~1996).
The distance was surprisingly small, i.e.~$17.1\pm1.7\,$Mpc, and was
precipitately interpreted as {\sl the} distance of the Virgo cluster (Mould
et al.~1995; Kennicutt, Freedman, \& Mould 1995), although already de 
Vaucouleurs (1982)
had derived an exceptionally small (relative) distance of this galaxy, and
in spite of the wide angular separation of the Virgo spiral members
and the correspondingly important depth effect. The next two spirals in the
Virgo complex with HST Cepheid distances did seemingly confirm the small
cluster distance (Saha et al.~1996a, 1996b). Yet all galaxies were selected
to be \lq\lq easy\rq\rq , i.e.~that their resolution had to be at least
\lq\lq good to fair\rq\rq ~(Sandage \& Bedke 1988), and there was therefore
an a priori expectation that they were nearer than average. (See also the
evidence from SN$\,$Ia models, Branch, Nugent, \& Fisher 1996.)

Indeed a fourth Virgo cluster spiral, NGC$\,$4639, which was selected
for observations with HST because it hat produced a supernova of type
Ia, and which is an unquestionable cluster member on the basis of its
{\sl low} redshift, gives a much larger Cepheid distance, i.e.~$25.1\pm
2.5\,$Mpc (Sandage et al.~1996). The four Virgo galaxies with Cepheid
distances are compiled in Table~\ref{tab:HSTVirgo}. Two of the
galaxies lie in subcluster A; the remaining ones in the southeastern W
cloud. The latter are therefore quite uncertain cluster members.

\begin{table}
\caption{Virgo cluster galaxies with HST Cepheid distances}
\label{tab:HSTVirgo}
\begin{center} \footnotesize
\begin{tabular}{lcclcl}
\hline
%\noalign{\smallskip}
 & $v_0\,$km s$^{-1}$ & subcluster & selection & $(m-M)^0$ & source \\
%\noalign{\smallskip}
\hline
%\noalign{\smallskip}
NGC$\,$4321 & 1464 & A & resolution & $31.04\pm 0.17$ & Ferrarese et al.~1996 \\
NGC$\,$4496A & 1568 & (W) & resolution & $31.13\pm 0.10$ & Saha et al.~1996b \\
NGC$\,$4536  & 1646 & (W) & resolution & $31.05\pm 0.15$ & Saha et al.~1996a \\
NGC$\,$4639  & \phantom{1}860  & A & well obs.~SN & $32.00\pm 0.23$ & Sandage et al.~1995 \\
\end{tabular}
\end{center}
\end{table}

It is clear that the four distances in Table~\ref{tab:HSTVirgo} cannot
be averaged. All one can conclude is that the cluster center lies
somewhere between 17 and 25 Mpc.  At least a dozen randomly selected
cluster members would be needed to define a reliable Cepheid distance
of the cluster as such. All what one can say {\sl with certainty is
that $(m-M)_{\rm Virgo} > 31.1$ and hence, from eq.~(2.2), $H_0 <
71\,$km s$^{-1}$ Mpc$^{-1}$!}

An auxiliary route to come to a Cepheid-based Virgo cluster distance at
present is to use the Leo group which, judging from the Cepheids in 
NGC$\,$3368 (M$\,$96; Tanvir et al.~1995) and NGC$\,$3351 (M95; Graham et 
al.~1996), lies at a mean distance of $(m-M)_{\rm Leo}=30.22\pm 0.12$
(increased by 0.05 mag for a zeropoint offset of HST photometry of relatively
bright stars following Saha et al.~1996). The relative distance between
the Leo group and the Virgo cluster can be estimated using different
distance indicators (Table~\ref{tab:LeoVirgo}).

\begin{table}
\noindent \caption{Relative distance moduli between the Leo group and
the Virgo cluster}
\label{tab:LeoVirgo}
\begin{center} \footnotesize
\begin{tabular}{lll}
%\noalign{\medskip}
\hline
%\noalign{\smallskip}
 Method & $\Delta (m-M)_{\rm Virgo - Leo}$ & Source \\
%\noalign{\smallskip}
\hline
%\noalign{\smallskip}
 Tully-Fischer & $1.35\pm 0.20$ & Federspiel et al.~(1996) \\
 Globular clusters & $1.47\pm 0.42$ & Harris (1990) \\
 D$_{\rm n}-\sigma$ & $0.97\pm 0.29$ & Faber et al.~(1989) \\
 Planetary nebulae & $1.15\pm 0.30$ & Bottinelli et al.~(1991) \\
 Velocities & $1.30\pm 0.30$ & Kraan-Korteweg (1986) \\
%\noalign{\smallskip}
\hline
%\noalign{\medskip}
 mean: & $1.25\pm 0.13$ & \\
\end{tabular}
\end{center}
\end{table}

With the distance of the Leo group and the modulus difference from
Table~\ref{tab:LeoVirgo} the distance modulus of the Virgo cluster
becomes
\begin{equation}
(m-M)^0_{\rm Virgo}=31.47 \pm 0.21 .
\end{equation}
For brevity we will refer to this value in the following as the \lq\lq Cepheid
distance of the Virgo cluster\rq\rq ; this is additionally supported by the
fact that the value lies midway between the directly determined Cepheid
distances of the Virgo cluster members in Table~\ref{tab:HSTVirgo}.

A first shot at $H_0$ is obtained by combining eqs.~(2.2) and (2.3), leading to
$H_0=60\pm 10\,$km s$^{-1}$ Mpc$^{-1}$ (external error).

\subsection{The Virgo Cluster Distance from 21$\,$cm Line Widths}

The 21$\,$cm line width-absolute magnitude relation (or Tully-Fisher
[TF] relation) has been applied many times to derive a Virgo cluster
distance with variable success. The principal difficulty is that a
{\sl complete} sample of cluster spirals is required to avoid the well
known \lq\lq Teerikorpi cluster bias\rq\rq ~(Teerikorpi 1987, 1990),
which systematically leads to an underestimate of the cluster distance
if only the brightest cluster members are used. Two previous studies
using complete samples selected from the Virgo Cluster Catalogue (VCC;
Binggeli, Sandage, \& Tammann 1985), gave a rather large distance in
nearly perfect mutual agreement, if the different adopted zero-point
calibrations of the TF relation are adjusted (Kraan-Korteweg, Cameron,
\& Tammann 1988; Fouqu\'e et al.~1990).

It has been suggested to use the inverse TF relation (i.e.~line widths $w$
versus absolute magnitude, the latter as independent variable to guard against
selection bias of magnitude-selected samples, which the complete Virgo sample
is not). While the direct method is affected by magnitude errors, the indirect
method is sensitive to errors of the line widths. In the Virgo sample the
magnitudes are quite good, and it seems that the line widths errors are more
important; this speaks here against the indirect method. The most severe
disadvantage of the indirect method as applied to clusters is, however, that
all members must be assumed to lie at the same distance and any depth effect,
which is so severe for the Virgo spirals, must be denied. This justifies
in the following that only the direct method is used (see also Giovanelli
1996b).

The method requires the galaxian magnitudes to be corrected for the 
inclination-depen\-dent internal absorption, which is notoriously difficult.
On the assumption that the internal absorption is significantly smaller in
the infrared it has been proposed to use $I$ or $H$ magnitudes for the
TF relation (e.g.~Aaronson et al.~1982). However, for an ensemble of stars
embedded in dust the absorption decreases with wavelength much slower than
with an $1/\lambda$-law. A systematic investigation of a complete sample
of 171 Virgo galaxies and 46 Fornax galaxies in $UBVRI$ has shown that
the accuracy of the TF distances of the two clusters is nearly the same
for all wavelengths (Schr\"oder 1995). The situation may be somewhat more
favourable for $H$ magnitudes, but they are available only for a small
fraction of the Virgo spirals.

The 171 Virgo and 46 Fornax spirals offer the possibility to study the
realistic wavelength dependence of the internal absorption by requiring
that the TF relation has a minimum scatter in each wavelength {\sl and}
that the distances of the individual cluster galaxies are independent on
average of the inclination. The result is that the internal absorptions
in $U$, $V$, $R$, and $I$ are 107, 91, 86 and 80\%, respectively, of the
absorption in $B$ (Schr\"oder 1995). The value for $A_I$ of 80\% is
independently confirmed by Giovanelli et al.~(1994). 

The calibration of the TF relation has been dramatically improved by the
advent of Cepheid distances form HST. There are now 14 Cepheid distances
available of spirals suitable for the TF calibration. Two late-type {\sl 
bona fide} companions of M$\,$101 can be added; the data are given
in Table~\ref{tab:TFcali}. The resulting calibration
in $B$ is shown in Fig.~2. For further details see Federspiel, Tammann \& 
Sandage (1996).

\begin{table}
\caption{Galaxies with Cepheid distances for the
calibration of the Tully-Fisher relation}
\label{tab:TFcali}
\begin{center} \footnotesize
\begin{tabular}{lcllccl}
\hline
%\noalign{\smallskip}
 Name & Hubble- &  $(m-M)$ & Source & $M_{\rm B}$  & 
 $i_{\rm RC3}$ & $\log w$ \\ 
      & type & \quad mag & & mag & $^{\circ}$ & \\
 (1) & (2) & \quad (3) & (4) & (5) & (6) & (7) \\
\noalign{\smallskip}
\hline
\noalign{\smallskip}
 N224 (M31) & 3 & 24.44 & Madore \& Freedman (1991)& -21.08& 78 & 2.728 \\ 
 N300 & 5 & 26.67 & Madore \& Freedman (1991) & -18.18 & 44 & 2.296 \\
 N598 (M33) & 5 & 24.63 & Madore \& Freedman (1991) & -18.90 & 55 & 2.357 \\
 N925 & 5 & 29.84$^{\ast}$ & Silbermann et al.~(1996a) & -19.80 & 58 & 
  2.382 \\ 
 N1365 & 4 & 31.30$^{\ast}$ & Silbermann et al.~(1996b) & -21.36 & 59 & 
  2.648 \\
 N2403 & 5 & 27.51 & Tammann \& Sandage (1968) & -19.13 & 62  & 2.415 \\
       &   &       & Madore \& Freedman (1991) &        &     &       \\
 N3031 (M81) & 3 & 27.80$^{\ast}$ & Freedman \& Madore (1994) & -20.46 & 65 & 
   2.667 \\
 N3351 (M95) & 3 & 30.01$^{\ast}$ & Graham et al.~(1996) & -19.77 & 50 & 
   2.538 \\
 N3368 (M96) & 2 & 30.32$^{\ast}$ & Tanvir et al.~(1995) & -20.47 & 50 & 
   2.649 \\
 N3621 & 5 & 29.85$^{\ast}$ & Macri et al.~(1996) & -19.55 & 57 & 2.509 \\
 N4321 (M100) & 5 & 31.04$^{\ast}$ & Ferrarese et al.~(1996) & -21.16 & 
  \phantom{0}36$\dagger$ & 2.725 \\
% N4496A & 5 & 31.13$^{\ast}$ & Saha et al.~(1996b)& -19.46 & 43 & 2.377 \\
 N4536 & 4 & 31.11$^{\ast}$ & Saha et al.~(1996a) & -20.49 & 66 & 2.548 \\
 N4639 & 3 & 32.00$^{\ast}$ & Sandage et al.~(1996)& -20.05 & 50 & 2.617 \\
 N5204 & 7 & 29.34& like M101 & -17.90 & 57 & 2.131 \\
 N5457 (M101) & 5 & 29.34$^{\ast}$ & Sandage \& Tammann (1974) & -21.03 & 
  \phantom{0}27$\dagger$ & 2.665 \\
       &      &   & Kelson et al.~(1995)  &   &  & \\
 N5585 & 7 & 29.34 & like M101 & -18.39 & 52 & 2.260 \\
%\noalign{\smallskip}
%\noalign{\smallskip}
\multicolumn{3}{l}{$^{\ast}$ Cepheid distance from HST} & & & & \\
\multicolumn{4}{l}{$\dagger$ the inclination is from a velocity map of the
galaxy} & & & \\
\end{tabular}
\end{center}
\end{table}

\begin{figure}
\begin{center}
\leavevmode
\epsfxsize 12.8cm
\epsffile{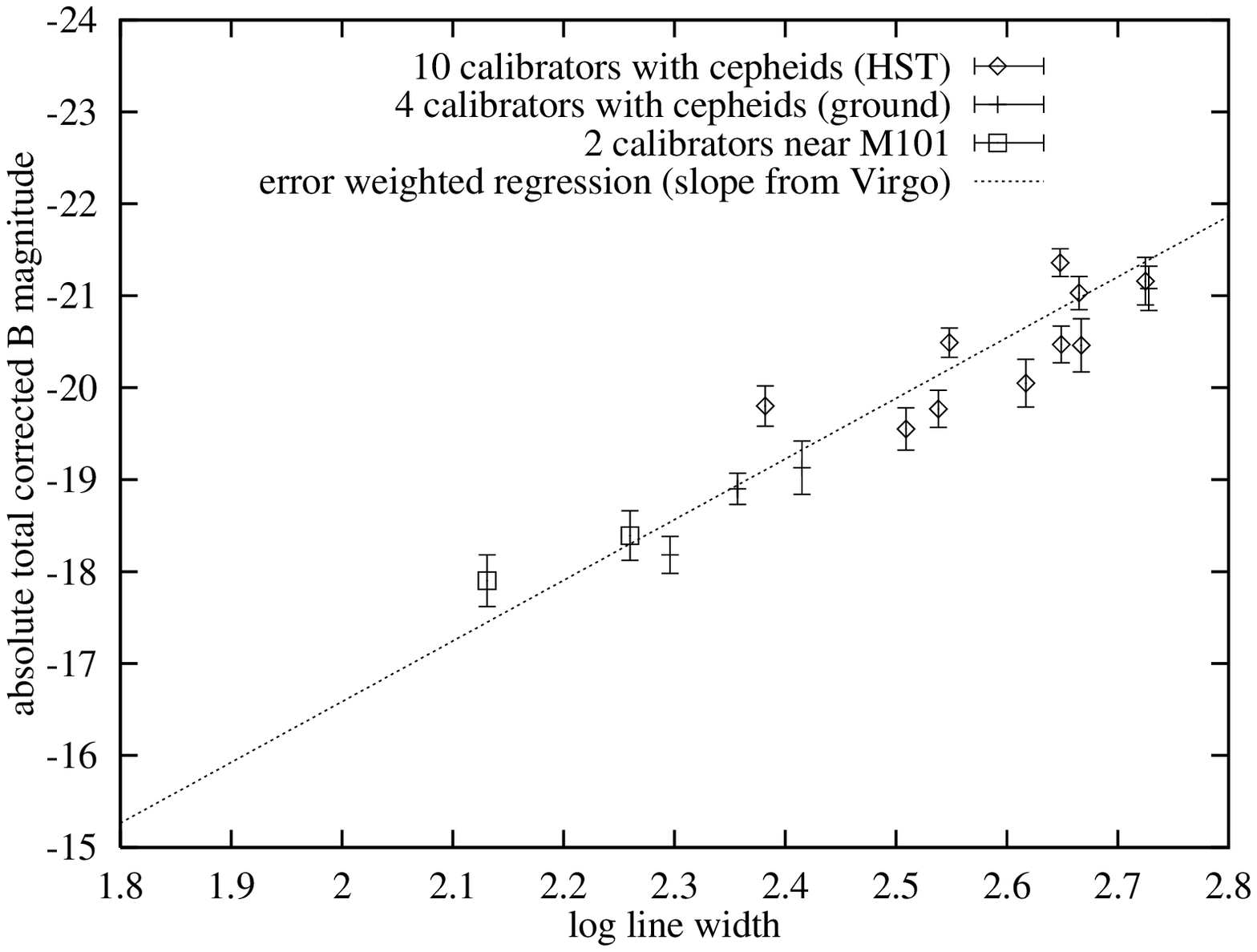}
%\vspace{10cm}
\caption{Tully-Fisher relation for the calibrators. Ten
calibrators have Cepheid distances determined with HST, 4 have ground-based
Cepheid distances, 2 are members of the relatively tight M$\,$101 group without
individual Cepheid distances. The error bars show the total errors in
absolute magnitude which were used as weights for the regression. The dotted
line represents the adopted calibration of eq.~(2.4)}
\end{center}
\end{figure}
A free linear regression to the calibrators yields a slope $M_{\rm
B}\propto -6.11 \log w$; but a better determined slope comes from the
complete sample of 49 inclined ($i>45^{\circ}$) spiral members of the
Virgo cluster [cf.~eq.~(2.5)].  With this slope the weighted
regression becomes
\begin{equation}
M_{\rm B}=-6.60 \log w - (3.38\pm 0.10)
\end{equation}
with a scatter of only $\sigma_{\rm M}=0.38$. [A larger sample of field
galaxies shows that the scatter actually depends strongly on the line
width $w$ (Federspiel, Sandage, \& Tammann 1994).

A fit of 49 Virgo cluster spirals gives 
\begin{equation}
m_{\rm B}= -6.60 \log w + (28.30\pm 0.08), \ \ \ \ \sigma=0.57.
\end{equation}
The larger scatter here is obviously due to the depth effect of the Virgo
cluster.

A combination of eqs.~(4) and (5) immediately gives the Virgo cluster
modulus of $(m-M)_{\rm Virgo}=31.68\pm 0.13$. However, this value depends
unexpectedly strongly on the choice of the input parameters. So far the
values of $m_{\rm B}$, internal absorption $A^i$, and Galactic absorption
$A^0$ are taken from the RC3 (de Vaucouleurs et al.~1991; $A^0$ actually
from Burstein \& Heiles 1984), and the line widths $w$ from LEDA (Lyon
Extragalactic Data Base). If these input parameters are taken from other
sources, one obtains distance moduli as given in Table~\ref{tab:TFinput}.

\begin{table}
\noindent \caption{The dependence of the Virgo TF modulus on input parameters}
\label{tab:TFinput}
\begin{center} \footnotesize
\begin{tabular}{clc}
%\noalign{\smallskip}
\hline
%\noalign{\smallskip}
solution & input parameters & $(m-M)_{\rm Virgo}$ \\ 
%\noalign{\smallskip}
\hline
%\noalign{\smallskip}
1 & $m_{\rm B}$, $A^i$, $A^0$ from RC3, $\log w$ from LEDA & $31.68\pm 0.13$ \\
2 & $m_{\rm B}$ from Schr\"oder 1995, others like solution 1 & $31.71\pm 
 0.13$ \\
3 & like solution 1, but $\log w$ from Huchtmeier \& Richter (1989) & 
 $31.70\pm 0.14$ \\
4 & like solution 1, but $A^i$ from RSA & $31.86\pm 0.13$ \\
5 & like solution 4, but $\log w$ from Huchtmeier \& Richter (1989) & 
 $31.88\pm 0.13$ \\
6 & like solution 1, but $A^0=0$ & $31.73\pm 0.13$ \\
%\noalign{\smallskip}
\hline
%\noalign{\smallskip}
mean 1-5 & & $31.77\pm 0.13$ \\
\end{tabular}
\end{center}
\end{table}

The result in Table~\ref{tab:TFinput} depends on the {\sl assumption}
that field and cluster galaxies obey the same TF relation. This
assumption can now be tested with the complete $UBVRI$ photometry of
Schr\"oder (1995). It turns out that the cluster galaxies are redder
in $(B-I)$ (corrected for internal and galactic reddening) at a given
line width $w$ than the 16 calibrators which are predominantly field
galaxies; the former are also H$\,$I-deficient.  The hydrogen
deficiency $D_{\rm H\,I}$ is here defined following Yasuda, Fukugita,
\& Okamura (1996). The colour residuals $\Delta (B-I)=(B-I)_{\rm obs.} -
(B-I)_{\rm calib.}$ at fixed line width correlate well with $D_{\rm
H\,I}$ (Schr\"oder \& Tammann 1996).

The consequence for the individual TF distances of the Virgo galaxies
is clear. In $U$, $B$, and $V$ cluster galaxies with small $\Delta
(B-I)$ give relatively small distances, while they give somewhat large
distances in $I$. For red galaxies [large $\Delta (B-I)$] the opposite
holds.  The individual distances in $R$ show the minimum dependence on
$\Delta (B-I)$.  Analoguously H$\,$I-normal galaxies give smaller
distances than H$\,$I-deficient ones in $U$, whereas the opposite is
true in $V$, $R$, and $I$. Here the smallest $D_{\rm H\,I}$ dependence
is found in the $B$-band.  The net effect is that the {\sl mean}
cluster distance decreases seemingly with increasing wavelength, the
$I$ modulus being 0.5 mag smaller than the $U$ modulus (cf.~column 2
of Table~\ref{tab:VirgoUBVRI}; Schr\"oder \& Tammann 1996).

From this it is clear that the systematic difference of the
calibrators and the cluster members in colour $(B-I)$ and in
H$\,$I-deficiency requires the cluster galaxies to be reduced to the
mean colour $(B-I)$ of the calibrators and to $D_{\rm H\,I}=0$. The
resulting distance moduli are shown in Table~\ref{tab:VirgoUBVRI}.

\begin{table}
\noindent \caption{Mean corrected TF Virgo moduli as a function of wavelength}
\label{tab:VirgoUBVRI}
\begin{center} \footnotesize
\begin{tabular}{cccc}
%\noalign{\smallskip}
\hline
%\noalign{\smallskip}
 passband & $(m-M)_{\rm Virgo}$ & $(m-M)_{\rm Virgo}$ & $(m-M)_{\rm Virgo}$ \\
          & uncorrected         & corrected for $\Delta (B-I)$ & corrected 
for $D_{\rm H\,I}$ \\
  (1) & (2) & (3) & (4) \\
%\noalign{\smallskip}
\hline
%\noalign{\smallskip}
U & 31.88 & 31.66 & 31.82 \\
B & 31.76 & 31.62 & 31.76 \\
V & 31.70 & 31.63 & 31.74 \\
R & 31.66 & 31.63 & 31.72 \\
I & 31.60 & 31.63 & 31.69 \\
%\noalign{\smallskip}
\hline
%\noalign{\smallskip}
mean & 31.72 & 31.63 & 31.75 \\
\end{tabular}
\end{center}
\end{table}

For the internal absorption it has been assumed here that
$A^i_{\lambda} = f_{\lambda} A^i_{\rm B}$ with $f_{\lambda}=1.07$,
1.00, 0.91, 0.86, and 0.80 for $U$, $B$, $V$, $R$, and $I$,
respectively (cf.~above). The moduli were calculated with five
different sets of input parameters, corresponding to the five
solutions in Table~\ref{tab:TFinput}. Only the {\sl mean} values of
the five solutions are shown in Table~\ref{tab:VirgoUBVRI}. [The
moduli in Schr\"oder \& Tammann (1996) were smaller by 0.15 mag
because the values of $A^i$ were 15\% higher (resulting in a decrease
of 0.05 mag) and because only the modulus corresponding to solution 1
was shown.] As can be seen in Table~\ref{tab:VirgoUBVRI} the
homogenization by colour of cluster galaxies and calibrators yields
consistent cluster moduli, whereas the homogenization by $D_{\rm
H\,I}$ leaves a residual drift with wavelength of 0.13 mag.

The best one can do is to average the mean corrected moduli columns
(3) and (4) in Table~\ref{tab:VirgoUBVRI} to obtain 
\begin{equation}
(m-M)_{\rm Virgo}=31.69\pm 0.15,
\end{equation}
where the error is the estimated internal error.

The general conclusion is that the application of the TF relation is
considerably more intricate than often realized. It not only takes
multicolour information for {\sl complete} cluster samples, but the result
is also sensitive to the input parameters. The Virgo modulus in $B$ is too
large by 0.07 mag, the one in $I$ too short by 0.09 mag. These values
may change from cluster to cluster depending on the colour excess and
the H$\,$I-deficiency of the spiral members.

\subsection{The Virgo Cluster Distance from SNe$\,$Ia}

Hamuy et al.~(1996) have determined the mean apparent peak magnitude of five
blue, particularly well observed SNe$\,$Ia in the Virgo cluster, obtaining
$m_{\rm B} ({\rm max})=12.16\pm 0.20$ mag and $m_{\rm V} ({\rm max})=12.07\pm 
0.20$ mag. In perfect agreement a somewhat larger sample of eight SNe$\,$Ia,
including objects
with older photometry, gives $m_{\rm B} ({\rm max})=12.10\pm 0.13$
mag and $m_{\rm V} ({\rm max})=12.11\pm 0.16$ mag (Sandage \& Tammann 1995).
Taking the larger sample, because it is less sensitive to depth effects of
the Virgo cluster, and combining it with the absolute calibration,
through Cepheids observed with HST, of $M_{\rm B} ({\rm max})=-19.53\pm 0.07$ 
mag and $M_{\rm V} ({\rm max})=-19.49\pm 0.07$ mag (Saha 1996), one
obtains an average cluster modulus of
\begin{equation}
(m-M)_{\rm Virgo}=31.62\pm 0.16 {\rm \ mag}.
\end{equation} 
The Cepheid-calibrated SNe$\,$Ia luminosities are in nearly perfect
agreement with the results of SN models. We will briefly return to the
latter in Section 5.

\subsection{Other Distances to the Virgo Cluster}
The peak of the luminosity function (LF) of {\sl globular clusters}
(GC) has frequently been used as a standard candle. A modern
calibration of the GCs in the Galaxy and in M31 combined with a
compilation of published GCLFs in five Virgo ellipticals has led to a
Virgo modulus of $(m-M)_{\rm Virgo}=31.75\pm 0.11$ (Sandage \& Tammann
1995). Meanwhile Whitmore et al.~(1995) found a very bright peak
magnitude in $V$ and $I$ for NGC~4486, which is well determined with
HST and which corresponds, with our precepts, to a modulus of
$31.41\pm 0.28$ (Sandage \& Tammann 1996). However, the GCs in
NGC~4486 have a bimodal colour distribution which is suggestive of age
differences and possible merger effects (Fritze-von Alvensleben 1995;
Elson \& Santiago 1996). Turning a blind eye to this problem and
averaging over all available GCLFs in Virgo we obtain
$(m-M)_{\rm Virgo}=31.67\pm 0.15$. We are aware that the method may still
face considerable problems.

Yet there comes some reassurement from a reversal of the argument.
The best Virgo modulus from five methods in Table 7 below (excluding
the globular clusters!) gives $(m-M)=31.66\pm 0.08$. The mean turnover
magnitude of the GCLF of eight {\sl E/S0 Virgo} cluster members gives $\langle
m_{\rm V}^0\rangle = 24.03\pm 0.08$ from a compilation of Whitmore
(1996). The resulting turnover magnitude of $M_{\rm V}^0 = -7.63\pm 0.11$
is in fortuitous agreement with the RR-Lyr and Cepheid-based calibration
of $M_{\rm V}^0 = -7.62\pm 0.08$ (Sandage \& Tammann 1995) from two
{\sl spiral} galaxies (Milky Way and M$\,$31). This is the first direct
evidence that the mean turnover luminosity of the GCLF is the same in
early-type galaxies and spirals to within the measuring accuracy. A point
of considerable worry is on the other hand that Whitmore (1996) finds
the turnover magnitude of seven {\sl early-type} Fornax galaxies to be 
0.23 mag {\sl brighter} than that of eight certain Virgo cluster members,
although at least the early-type Fornax members are certainly more
distant (cf.~Sect.~3).

The $D_n-\sigma$ {\sl method}, normally applied to ellipticals, was extended
to the bulges of S0 and spiral galaxies by Dressler (1987). Using the bulges
of the Galaxy, M31, and M81 as local calibrators, one obtains $(m-M)_{\rm
Virgo} =31.85\pm 0.19$ (Tammann 1988).

{\sl Novae} are potentially powerful distance indicators through their
luminosity-decline rate relation. Using the Galactic calibration of
Cohen (1985), Capaccioli et al.~(1989) have found the apparent
distance modulus of M31 to be $(m-M)_{\rm AB}=24.58\pm 0.20$
(i.e. somewhat less than indicated by Cepheids). From six novae in
three Virgo ellipticals Pritchet \& van den Bergh (1987) concluded
that the cluster is more distant by $7.0\pm 0.4$~mag than the apparent
modulus of M31, implying $(m-M)_{\rm Virgo}=31.58\pm 0.45$. 
The result carries still low weight, but is interesting because it is
based on novae exclusively. HST observations, although
time-consuming, of novae in the Virgo cluster could much improve this
{\sl independent} result.

For some time it has been thought that the luminosity function of the
shells of {\sl planetary nebulae} (PNe) in the light of the 
$\lambda 5007\,$\AA ~emission line had a sharp universal cutoff at 
$M_{5007}=-4.48$ mag. On the basis of this assumption a very small Virgo
cluster modulus of $(m-M)=30.84$ was suggested (Jacoby, Ciardullo \& Ford 
1990). However, it was pointed out that the luminosity of the brightest PN
shells depends on the population size, i.e.~on the luminosity of the
parent galaxy, in agreement with statistical expectations (Bottinelli et
al.~1991). In fact it was shown that the available PN data agree perfectly
well with a Virgo cluster modulus of $(m-M)\sim 31.6$ if allowance is
made for the population size effect (Tammann 1993). Numerically simulated
luminosity functions of the shell luminosities have since confirmed that
the peak luminosities depend on the sample size {\sl and} on the population
age (M\'endez et al.~1993). It has therefore been proposed to use
essentially the {\sl shape} of the $\lambda 5007\,$\AA ~luminosity function
(Soffner et al.~1996), but no results of this new method are available
yet for the Virgo cluster.

{\sl Surface brightness fluctuations} (SBF) of ellipticals and of the
bulges of spirals have also been proposed as distance indicators
(Tonry \& Schneider 1988). The first \lq\lq test\rq\rq ~has remained
rather unconvincing, spreading the {\sl elliptical} Virgo cluster
members over an interval of 12 to 24 Mpc (Tonry, Ajhar \& Luppino
1990); this interval was interpreted as real although early-type
galaxies are known to be concentrated in the {\sl cores} of galaxy
clusters.  Moreover the individual distances correlate with the
Mg$_{2}$ index (Lorenz et al.~1993). A re-evaluation of the method has
remedied these objections (Tonry et al.~1997), but some of the results
remain unconvincing.  For instance the modulus distance of the Virgo
cluster and the Leo group is suggested to be $0.89\pm 0.08$ mag as
compared to $1.25\pm 0.13$ mag from Table~\ref{tab:LeoVirgo}. More
telling is still the impossible SFB distance of the Virgo cluster of
$(m-M)=31.03\pm 0.05$ (!). This is not only $0.66\pm 0.09$ mag less
than from all other evidence (cf.~Table~\ref{tab:Virgomethods} below),
but it would also imply a mean absolute magnitude of $M_{\rm B} ({\rm
max})=-18.93\pm 0.14$ mag of the eight standard type Ia SNe of the
Virgo cluster, which is excluded by the seven Cepheid-calibrated
SNe$\,$Ia giving $M_{\rm B} ({\rm max})=-19.53\pm 0.07$ (Saha 1996)
and by all existing type Ia models (cf.~Section 4). The conclusion
that the SFB distances beyond $(m-M)\sim 30.0$ are too small by $\ge
0.5$ mag will soon be decisively tested by NGC$\,$7331, for which the
SFB method gives a quite small distance of $(m-M)=30.39\pm 0.10$
(Tonry et al.~1997), and for which a Cepheid distance will soon become
available from HST (Hughes 1996).

\subsection{The adopted Virgo cluster distance and the value of
$H_0$ thereof}

The valid evidence for the Virgo cluster distance is compiled
in Table~\ref{tab:Virgomethods} from the preceding paragraphs. The six
methods give very consistent results. This is remarkable in two respects.
First, the methods include independent distance scales: the Cepheids,
TF-distances, and SNe$\,$Ia depend on the zero point of the P-L relation
of Cepheids, the $D_{\rm n}-\sigma$ method does not only depend on the
Cepheid distances of M$\,$31 and M$\,$81, but also on the independent
size of the Galactic bulge, the globular clusters rest on the P-L relation
of RR Lyrae stars, and the novae rely on Cohen's (1985) Galactic calibration.
Secondly the different distance determinations comprise spiral and E/S0
galaxies. This suggest that these two types of galaxies center about a
common distance.

The best Virgo cluster modulus of $(m-M)_{\rm Virgo}=31.66\pm 0.08$
from Table~\ref{tab:Virgomethods} inserted into eq.~(2.2) gives the
large-scale value (out to at least $\sim 10\,000\,$km s$^{-1}$) of
\begin{equation}
H_0= 54\pm 4.
\end{equation}

\begin{table}
\noindent \caption{The Virgo cluster modulus from various methods}
\label{tab:Virgomethods}
\begin{center} \footnotesize
\begin{tabular}{lcc}
%\noalign{\smallskip}
\hline
%\noalign{\smallskip}
 method & $(m-M)_{\rm Virgo}$ & Hubble type \\ 
%\noalign{\smallskip}
\hline
%\noalign{\smallskip}
Cepheids & $31.47\pm 0.21$ & S \\
TF & $31.69\pm 0.15$ & S \\
SNe$\,$Ia & $31.62\pm 0.16$ & S \\
Globular Cl. & $31.67\pm 0.15$ & E \\
$D_{\rm n}-\sigma$ & $31.85\pm 0.19$ & E, S0 \\
Novae & $31.46\pm 0.40$ & E \\
%\noalign{\smallskip}
\hline
%\noalign{\smallskip}
mean & $31.66\pm 0.08$ & \\
\end{tabular}
\end{center}
\end{table}

\section{$H_0$ from the Fornax or Coma clusters?}

The Fornax cluster has in spite of being an irregular cluster
a large fraction of early-type galaxies. The distribution in the sky
of E/S0 and dE galaxies is shown in Fig.~6a, that of spiral and Im members
in Fig.~6b. The relevant data are taken from Ferguson (1989). The well known
Hubble type-density relation holds also here: The early-type galaxies
are clearly more concentrated toward the cluster center; the spirals
may still be more widely distributed as shown in Fig.~6b because the
survey limits impose an artificial cutoff. The cluster diameter traced
by spirals is $\sim 8^{\circ}$, which corresponds -- assuming sphericity --
to a depth effect of 0.25 mag.

\begin{figure}
\begin{center}
\leavevmode
\epsfxsize 7.cm
\epsffile{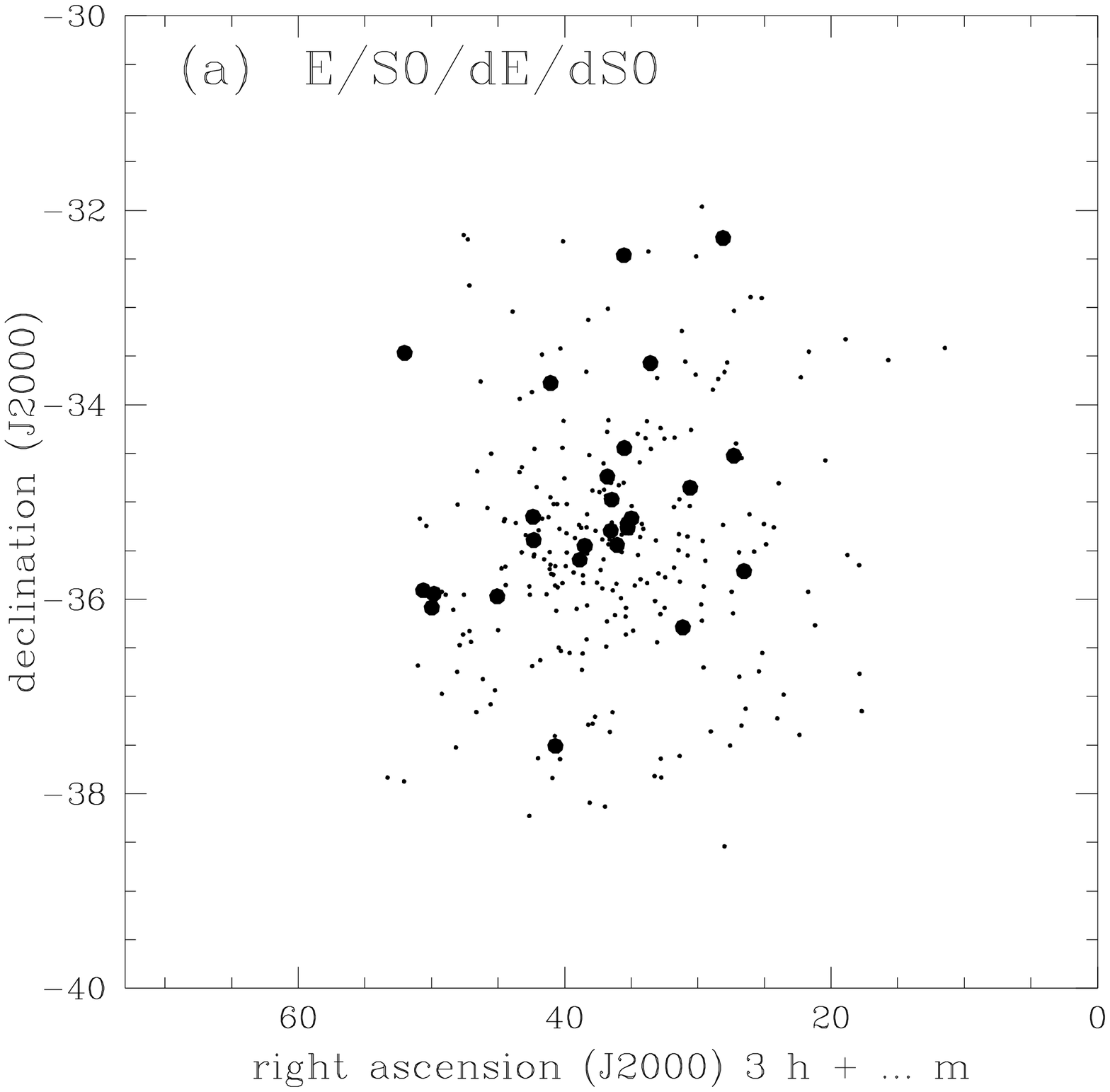}
\epsfxsize 7.cm
\epsffile{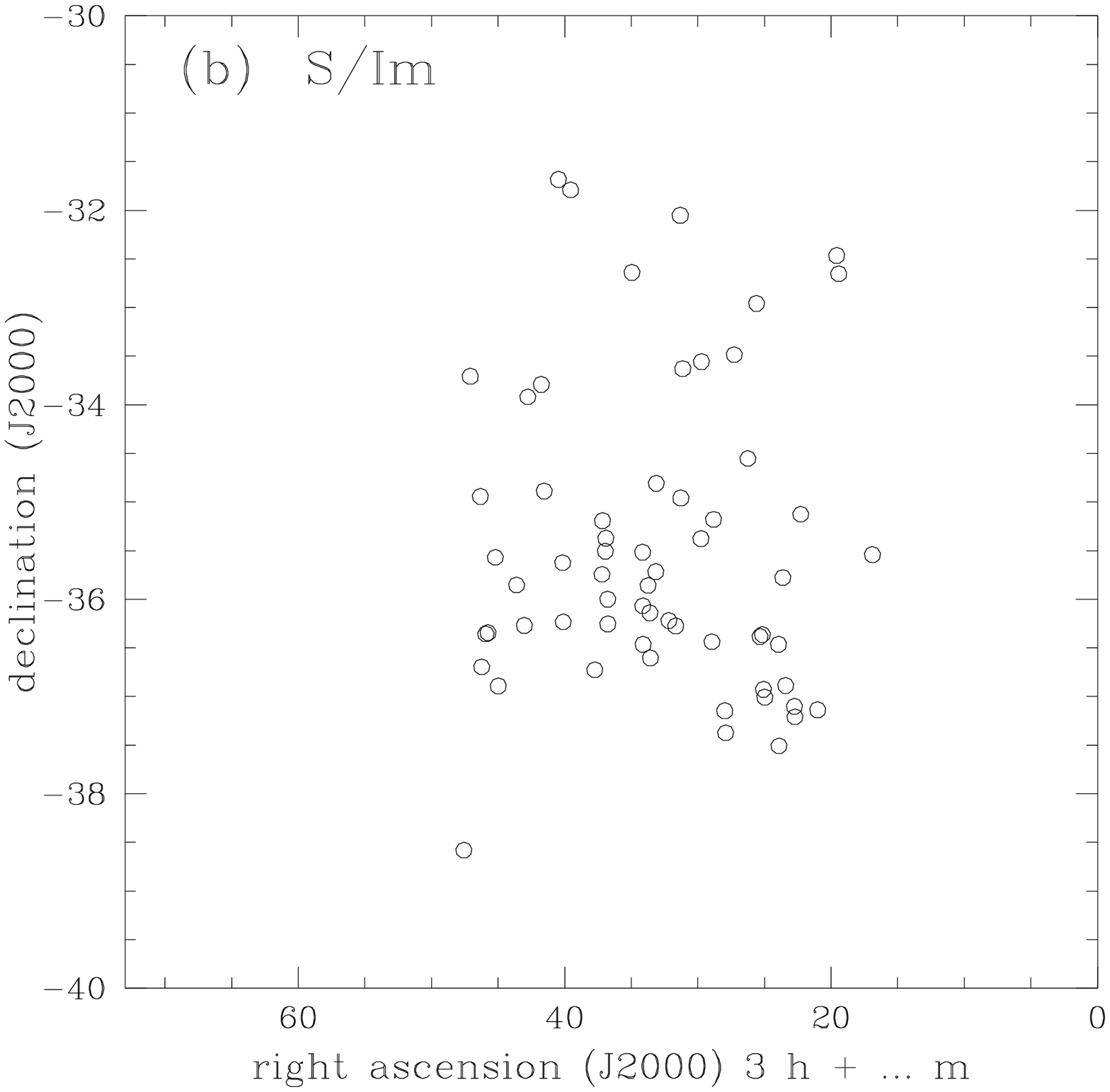}
%\vspace{10cm}
\caption{Apparent distribution of Fornax cluster galaxies in the sky:
the E/S0/dE members are clearly concentrated towards the cluster center (a),
whereas the spirals and irregulars are more widely distributed (b)}
\end{center}
\end{figure}

The distance to the Fornax cluster is notoriously poorly known. Even
for the relative distance between Virgo and Fornax amazingly different
values have been published. Such relative distances from the
literature of the last 30 years are compiled in
Table~\ref{tab:ForVirgo}. The values are separated here for early-type
galaxies and S/Im galaxies. The result is that the latter have smaller
relative distance moduli than the E/S0/dE members by $0.35\pm 0.09$
mag. This is not due to a distance difference within the Virgo
cluster, because here the two types of galaxies seem to be closely at
the same mean distance. It would be premature to take the separation
in space between the early-type and late-type Fornax members at face
value, but it should be taken as a clear warning that the Fornax
cluster may be elongated along the line of sight. 

\begin{table}
\noindent \caption{Relative distances between the Fornax and Virgo clusters}
\label{tab:ForVirgo}
\begin{center} \footnotesize
\begin{tabular}{lr@{$\ \ \pm$}lr@{$\ \ \pm$}ll}
%\noalign{\smallskip}
\hline
%\noalign{\smallskip}
 method & \multicolumn{2}{c}{$\Delta (m-M)$} & 
 \multicolumn{2}{c}{$\Delta (m-M)$} & reference \\
 & \multicolumn{2}{c}{Fornax-Virgo} & \multicolumn{2}{c}{Fornax-Virgo} & \\  
 & \multicolumn{2}{c}{spirals} & \multicolumn{2}{c}{early types} & \\
%\noalign{\smallskip}
\hline
%\noalign{\smallskip}
TF in $B$ & $-0.40$ & $0.10$ & \multicolumn{2}{c}{} & Schr\"oder (1995) \\
TF in $I$ & $-0.06$ & $0.15$ & \multicolumn{2}{c}{} & Bureau et al.~(1996) \\
SFB  & \multicolumn{2}{c}{} & $0.20$ & $0.08$ & Tonry et al.~(1997) \\ 
globular clusters  & \multicolumn{2}{c}{} & $0.08$ & $0.09$ & Kohle et al.~(1996) \\ 
surf.$\,$brightn.-lum.$\,$rel.~of dEs & \multicolumn{2}{c}{} & $0.40$ & $0.12$ & Jerjen (1995)\\
SN$\,$Ia & \multicolumn{2}{c}{} & $0.36$ & $0.12$ & Sandage \& Tammann (1995) \\
PNe & \multicolumn{2}{c}{} & $0.24$ & $0.10$ & Mc Millan et al.~(1993) \\
SN$\,$Ia & \multicolumn{2}{c}{} & $0.09$ & $0.14$ & Hamuy et al.~(1991) \\
SFB & \multicolumn{2}{c}{} & $-0.16$ & $0.13$ & Tonry (1991) \\
globular clusters & \multicolumn{2}{c}{} & $-0.11$ & $0.20$ & 
Bridges et al.~(1991) \\
globular clusters & \multicolumn{2}{c}{} & $-0.5$ & $0.2$ & Geisler \& Forte (1990) \\
surf.$\,$brightn.-lum.$\,$rel.~of dEs & \multicolumn{2}{c}{} & $-0.19$ & $0.15$ & Bothun et al.~(1989) \\
lum.$\,$vel.$\,$dispersion & \multicolumn{2}{c}{} & $-0.48$ & $0.58$ & Pierce (1989) \\
lum.$\,$vel.$\,$-surf.$\,$brightn.$\,$rel. & \multicolumn{2}{c}{} & $-0.14$ & $0.43$ & Pierce (1989) \\
TF in $H$ & $-0.25$ & $0.23$ & \multicolumn{2}{c}{} & Aaronson et al.~(1989) \\
$D_{\rm n}-\sigma$ & \multicolumn{2}{c}{} & $0.14$ & $0.17$ & Faber et al.~(1989) \\
surf.$\,$brightn.-lum.$\,$rel.~of E/dEs & \multicolumn{2}{c}{} & $-0.02$ & $0.20$ & Ferguson \& Sandage (1988a) \\
surf.$\,$brightn.-lum.$\,$rel.~of dEs & \multicolumn{2}{c}{} & $-0.5$ & $0.2$ & Caldwell \& Bothun (1987) \\
$D_{\rm n}-\sigma$ (Mg$_2$) relation & \multicolumn{2}{c}{} & $0.21$ & $(0.2)$ & Dressler (1987)\\
TF in $H$ & $-0.2$ & $0.25$ & \multicolumn{2}{c}{} & Aaronson \& Mould 1983 \\
TF in $V,r,IV$ & $-0.20$ & $0.18$ & \multicolumn{2}{c}{} & Visvanathan (1983) \\
colour-lum.$\,$relation & \multicolumn{2}{c}{} & $0.23$ & $0.20$ & Griersmith (1982) \\
revision Visvanathan \& & \multicolumn{2}{c}{} & \multicolumn{2}{c}{} & \\
Sandage (1977) & \multicolumn{2}{c}{} & $0.16$ & $0.17$ & Aaronson et al.~(1980) \\
globular clusters & \multicolumn{2}{c}{} & $0.70$ & $0.3$ & de Vaucouleurs (1977) \\
colour-lum.$\,$relation & \multicolumn{2}{c}{} & $0.32$ & $0.23$ & Visvanathan \& Sandage (1977)\\
brightest cluster galaxies & \multicolumn{2}{c}{} & $0.4$ & $0.3$ & Dawe \& Dickens (1976) \\
brightest cluster galaxies & \multicolumn{2}{c}{} & $0.36$ & $0.25$ & Sandage (1975) \\
Magn.$\,$\& diam.$\,$of brightest gal. & \multicolumn{2}{c}{} & $0.55$ & $0.3$ & de Vaucouleurs (1975) \\
brightest cluster galaxies & \multicolumn{2}{c}{} & $0.62$ & $0.25$ & Sandage \& Hardy (1973) \\
\hline
unweighted mean & $-0.22$ & $0.05$ & $0.13$ & $0.07$ \\
\end{tabular}
\end{center}
\end{table}

Velocities do not help to elaborate on the spatial structure of the Fornax
cluster. The mean velocity of 41 E/S0/dE galaxies is $v_{220}=1323\pm 48\,$km
s$^{-1}$ with a dispersion of $\sigma=307\,$km s$^{-1}$; the mean velocity
of 27 S/Im galaxies is $v_{220}=1436\pm 66\,$km s$^{-1}$ with an only 
slightly larger dispersion of $\sigma=343\,$km s$^{-1}$. The statistical
agreement of the mean velocities can be interpreted as the early- and late-type
members being at the same distance, but it could also be the result of the
late-type members lying in the foreground and falling away towards the
core of the cluster formed by early-type galaxies.

The mean velocity over all types is $v_{\rm LG}=1366\pm 50\,$km
s$^{-1}$ or $v_{220}=1338\,$km s$^{-1}$. The correction $\Delta
v_{220}$ for a self-consistent Virgocentric infall model is small
because the Fornax cluster lies far away from the Virgo cluster at an
angle of $133^{\circ}$ from the latter. But that means also that the
Fornax cluster may have its own peculiar motion of say $\pm 350\,$km
s$^{-1}$ (see Section 2.1). -- The cluster velocity in the CMB frame
can be inferred from eq.~(2.1) to be $v^{\rm CMB}=1300\pm 75\,$km
s$^{-1}$, but this depends entirely on its distance relative to the
Virgo cluster as given in Table~\ref{tab:clusters}. The reliability of
the latter value may be questioned in view of the data in
Table~\ref{tab:ForVirgo}.

The conclusion from this is that the Fornax cluster is much less suited
for the determination of $H_0$ than the Virgo cluster. The possible
spatial separation of the Fornax members, its expected non-negligible
peculiar motion, and the low weight of its $v^{\rm CMB}$ velocity
call for great caution.  

Three exercises to derive $H_0$ from the Fornax cluster may illustrate
the difficulties. Here we do not carry through the errors of $H_0$ caused
by the uncertain cluster velocity; the errors shown reflect only the errors
introduced by the distance errors.

(1) A Cepheid distance of $18.2\pm 1.3\,$Mpc has been published for
NGC 1365 (Silbermann et al.~1996), an exceptionally large spiral
galaxy in the Fornax field. If this value is confirmed and if it is
taken -- quite naively -- as {\sl the} distance of the cluster, one
obtains with an adopted cluster velocity of $v_{\rm LG}=1366\,$km
s$^{-1}$ $H_0=75\pm 6$, or with the mean velocity of only the spirals
($1436\,$km s$^{-1}$) $H_0=79\pm 6$.

(2) If one combines the modulus difference from
Table~\ref{tab:clusters} of $(m-M)_{\rm Fornax}-(m-M)_{\rm
Virgo}=0.23\pm 0.20$ with the best modulus of the Virgo cluster in
Table~\ref{tab:Virgomethods}, one finds $(m-M)_{\rm Fornax}=31.89\pm
0.22$.  This with $v_{220}=1338\,$km s$^{-1}$ gives $H_0=56\pm 6$.

(3) A yet better distance can be determined from SN 1992A, a normal
SN$\,$Ia in NGC$\,$1380. The cluster membership of this {\sl
early-type} (S0/a) galaxy, which lies close to the cluster core,
cannot seriously be questioned. The relevant data for SN 1992A are
given in Table~\ref{tab:SNFor} together with the data of SNe$\,$Ia
1980N and 1981D, which both occurred in the outlying Sa galaxy
NGC$\,$1316. The cluster membership of this galaxy is not certain, but
the data for all {\sl three} SNe$\,$Ia are so similar, that the
inclusion of NGC$\,$1316 only decreases, if anything, the inferred 
cluster distance. 

\begin{table}
\noindent \caption{Data for the three SN$\,$Ia in Fornax cluster galaxies}
\label{tab:SNFor}
\begin{center} \footnotesize
\begin{tabular}{llccc}
%\noalign{\smallskip}
\hline
%\noalign{\smallskip}
SN & galaxy & $m_{\rm B} ({\rm max})$ & $m_{\rm V} ({\rm max})$ & $(m-M)$ \\
(1) & (2) & (3) & (4) & (5) \\ 
%\noalign{\smallskip}
\hline
%\noalign{\smallskip}
1980N & NGC$\,$1316 & 12.49 & 12.44 & 31.75 \\
1981D & NGC$\,$1316 & 12.59 & 12.40 & 31.78 \\
1992A & NGC$\,$1380 & 12.60 & 12.55 & 31.86 \\
\hline
mean & & & & 31.80 \\
\end{tabular}
\end{center}
\end{table}

The apparent magnitudes of the three SNe$\,$Ia in columns (3) and (4)
in Table~\ref{tab:SNFor} are taken from Hamuy et al.~(1996). The
distance moduli in column (5) are the mean of the moduli in $B$ and
$V$. They are based on assumed absolute maximum magnitudes of $M_{\rm
B} ({\rm max})=-19.31$ and $M_{\rm V} ({\rm max})=-19.26$. These
absolute magnitudes are 0.20 mag {\sl fainter} than the mean of seven
SNe$\,$Ia calibrated with Cepheids (Saha et al.~1997).  The fainter
luminosities are indicated here because the calibrating SNe$\,$Ia lie
predominantly in late-type spirals and since SNe$\,$Ia in early-type
galaxies are generally somewhat fainter (Suntzeff 1996).  The
resulting mean Fornax cluster distance of $(m-M)=31.80\pm 0.20$ leads
together with $v^{\rm CMB}=1300\,$km s$^{-1}$ to $H_0=57\pm 6$.

The conclusion of point (1) to (3) is that the main body of the
Fornax cluster formed by early-type galaxies may well comply with
$H_0\approx 55$. If the small Cepheid distance of the large spiral
NGC$\,$1365 is correct, it lies certainly in the foreground, possibly
together with many other spirals of the field as suggested by the data
in Table~\ref{tab:ForVirgo}. In any case the Fornax cluster, whose
velocity in the CMB frame is in addition quite uncertain, cannot
provide at present a meaningful value of $H_0$.

It is sometimes suggested to use the Coma cluster to derive $H_0$ by
adding the relative distance modulus $\Delta (m-M)_{\rm Coma-Virgo}$
to the Virgo cluster modulus $(m-M)_{\rm Virgo}$ and by assuming
-- without justification -- that the peculiar motion of the Coma cluster
is negligible. In view of the {\sl many} cluster distances relative to
the Virgo cluster and of their Hubble diagram in Fig.~1 this is now a
step backwards. However, there is {\sl one} independent distance information
for the Coma cluster. From HST observations of the globular clusters
of NGC$\,$4881 Baum et al.~(1995) found a minimum distance of $r_{\rm Coma}
> 108\pm 11\,$Mpc. Since the depth effect of the cluster becomes
vanishingly small at the distance of Coma, this result holds for the
whole cluster. With $\Delta (m-M)_{\rm Coma-Virgo} = 3.80$ from Table 1
and from eq.~(2.1) one finds a cosmic velocity of $v_{\rm Coma}^{\rm CMB}=
6730\,$km s$^{-1}$ which gives, when combined with Baum's et al.~(1995)
lower distance limit, $H_0 < 62\pm 7$. (Note that a smaller value of
$\Delta (m-M)_{\rm Coma-Virgo}$, as preferred by some authors, gives a
lower value of $v_{\rm Coma}^{\rm CMB}$ and hence of $H_0$.

\section{$H_0$ from field galaxies}

Catalogues of field galaxies are flux-limited. This has the consequence,
in the presence of intrinsic luminosity scatter, that the mean luminosity
of the galaxies increases with distance. This does not only hold for
galaxies in general, but also for galaxies of a fixed rotation velocity,
velocity dispersion, colour etc.~-- whereever the dispersion is non-negligible.
The unfailing signature of Malmquist bias is -- if uncorrected -- that
it always leads to values of $H_0$ increasing with distance, contrary to
overwhelming evidence (for a discussion cf.~Sandage 1995). Uncorrected
distances of field galaxies (de Vaucouleurs 1979; Aaronson et al.~1986),
which were said to require high values of $H_0$, lead indeed to an 
unrealistic increase of $H_0$ with distance (Tammann 1987). 

A simple-minded compensation for the Malmquist bias has been to take
{\sl distant} galaxies from a catalogue with a fainter flux limit than
that of nearer ones (Sandage \& Tammann 1975). The next step was to
introduce a correction term to the mean absolute magnitudes which
increases linearly with distance in the sense to make distant galxies
brighter (San\-dage, Tammann \& Yahil 1979). Sophisticated Malmquist
corrections to TF distances were applied by Bottinelli et al.~(1986),
Lynden-Bell et al.~(1988), Sandage (1988a), Federspiel, Sandage \&
Tammann (1994), Giovanelli (1996b), Theureau et al.~(1996) and
others. In a case study Hendry \& Simmons (1990) have considered 184
Sc galaxies of luminosity class I and derived a \lq\lq naive\rq\rq
~value of $H_0=86$ (using an arbitrary zeropoint) if $\sigma_{\rm M}$
were zero; however, allowing for an internal dispersion of
$\sigma_{\rm M}=0.5$ and 1.0 mag they found a maximum likelyhood
solution of $H_0=56$ and $H_0=44$, respectively.

A representative sample of Malmquist-corrected $H_0$ determinations from
field galaxies with typically $v<4000\,$km s$^{-1}$ is compiled in 
Table~\ref{tab:HubMalm}.

\begin{table}
\noindent \caption{$H_0$ determinations from field galaxies corrected for
Malmquist bias}
\label{tab:HubMalm}
\begin{center} \footnotesize
\begin{tabular}{lll}
%\noalign{\smallskip}
\hline
%\noalign{\smallskip}
method& $H_0$ & source \\ 
%\noalign{\smallskip}
\hline
%\noalign{\smallskip}
Tully-Fisher, distance-limited, & & Richter \& Huchtmeier (1984); \\
\phantom{\quad}($v_{\rm LG}\le 500\,$km s$^{-1}$) & $50\pm 5$ & Sandage (1988a, 1994) \\
Tully-Fisher, flux-limited (distant) & $<60$ & Sandage (1994) \\
M$\,$101 look-alike diameters & $43\pm 11$ & Sandage (1993a) \\
M$\,$31 look-alike diameters & $45\pm 12$ & Sandage (1993b) \\
luminosity classes of spirals & $56\pm 5$ & Sandage (1996a) \\
M$\,$31, M$\,$101 look-alike luminosity & $55\pm 5$ & Sandage (1996a) \\
Tully-Fisher (using magn.+diam.) & $55\pm 5$ & Theureau et al.~(1996) \\
\hline
weighted mean & $53\pm 3$ & \\
\end{tabular}
\end{center}
\end{table}

It remains to point to an unexpected paradox. The 15 or so galaxies
outside the Local Group with known Cepheid distances give, in combination with
their velocities $v_{\rm LG}$ or $v_{220}$, a {\sl very local} value
of $H_0\approx 70\pm 5$ (cf.~Giovanelli 1996b), i.e.~$H_0$ seems to be
larger within the Virgo cluster circle than outside. This is contrary
to all previous (mainly Malmquist-generated) claims of $H_0$ increasing
outwards, but also contrary to physical expectations in view of the local
overdensity. Judgement about this result must await a still larger sample
of Cepheid distances. The result is questioned by the TF distances of
a {\sl complete} sample of spiral galaxies with $v_{\rm LG}< 500\,$km
s$^{-1}$ (Kraan-Korteweg \& Tammann 1979) giving $H_0=50\pm 5$ 
(cf.~Table~\ref{tab:HubMalm}).

\section{$H_0$ from geometric and physical methods}

The accurate geometric distance determination of LMC via the ring
of SN$\,$1987A (Panagia et al.~1996) is very important for the
extragalactic distance scale because it supports the Cepheid distance of
LMC of $(m-M)^0=18.50$ to within 0.1 mag. 

Other geometric distances come from VLBI observations of the radio remnants
of SNe$\,$II. They have provided two distances so far, viz.~of the nearby
galaxy M$\,$81 (Marcaide et al.~1995) and the Virgo galaxy NGC$\,$4321
(M$\,$100; Bartel 1991). The accuracy, however, does not match that of
the Cepheids.

There are plans to measure proper motions in external galaxies by means
of masers (e.g.~Elitzur 1992), but much progress is still needed to derive
distances which are directly useful for $H_0$. The velocity drift of
water maser lines close to the very massive nucleus of NGC$\,$4258 are
attributed to the centripetal acceleration of gravity; this together with
VLBI observations provides a distance of this galaxy of $6.4\pm 0.9\,$Mpc
(Miyoshi et al.~1995). -- Sparks (1994) has proposed to determine the
locus of maximum polarization of the scattered light echos of SNe; HST
observations for this project are under way.
 
Important results are coming from theoretical models of SNe. From
expanding-photo\-sphere models of SNe$\,$II Schmidt et al.~(1994) have
argued that $H_0=73\pm 6$, but their crucial dilution factor
(i.e.~deviation from black-body radiation) has been criticized by
Baron et al.~(1995; 1996) who argue for $H_0< 50$. There is less
controversy for SNe$\,$Ia. H\"oflich et al.~(1996) have derived
spectrum-fitting expanding-atmosphere luminosities for five SNe$\,$Ia
for which Cepheid distances are available. Their mean absolute
magnitudes of $M_{\rm B} ({\rm max})=-19.40\pm 0.20$ and $M_{\rm V}
({\rm max})=-19.37\pm 0.18$ are fainter by only $\Delta M_{\rm B}=
0.05\pm 0.21$ and $\Delta M_{\rm V}=0.10\pm 0.19$ than Saha's (1996)
Cepheid-calibrated mean luminosities, the only significant difference
being that H\"oflich's et al.~(1996) luminosities have larger scatter
than actually observed. The five theoretical models combined with the
Hubble diagram of SNe$\,$Ia beyond $v=1000\,$km s$^{-1}$ gives
$H_0=60\pm 6$.  Branch et al.~(1996) find an expanding-photosphere of
$M_{\rm B} ({\rm max})=-19.49\pm 0.35$ and $M_{\rm V} ({\rm
max})=-19.42\pm 0.35$ for SN$\,$Ia 1981B and from its $^{56}$Ni mass
the independent value $M_{\rm B} ({\rm max})\approx M_{\rm V} ({\rm
max})\approx-19.4\pm 0.2$. This should be compared with the marginally
fainter calibration from Cepheids (Saha 1996, Table 1). Ruiz-Lapuente
(1996) has derived absolute magnitudes from the late nebular spectra
of two SNe$\,$Ia (SN 1989B and SN 1990N) which are marginally fainter
by $0.26\pm 0.25$ mag than their Cepheid-calibrated luminosities.
 
The Sunyaev-Zeldovich effect (SZ) provides distances of X-ray clusters the
quality of which depends on a detailed gas model and a number of assumptions,
one of which being the sphericity of the gas volume. Several distance
determinations of the cluster A$\,$2218 yield $H_0=45\pm 20$ (McHardy et 
al.~1990; Birkinshaw \& Hughes 1994; Jones 1994; Lasenby \& Hancock 1995).
The unweighted mean of five other clusters from various authors (listed by
Raphaeli 1995) and of the Coma cluster (Herbig, Lawrence \& Readhead 1995)
is $H_0=60\pm 15$. Lasenby (1996) derives from two clusters, A$\,$2218 and
A$\,$1413, $H_0=42^{+12}_{-9}$. Raphaeli (1995) concludes in his review that
the method will be become competitive through a compilation of high-sensitivity
X-ray and SZ data on a {\sl large} cluster sample.

Distances of gravitationally lensed double quasars can also be determined
{\sl if} the time delay between the variablity of the two images and the
mass distribution of the deflector lens are known (Refsdal 1964). These
distant objects do not yield, however, the {\sl present} value of $H$.
One has therefore to {\sl assume} some value of $q_0$ to obtain $H_0$,
but the uncertainty is only of the order of $\sim 10$\%. The double quasar
QSO 0957+561 with a now well-determined time delay of $\tau=420$ days 
(Pelt et al.~1996; Thomson \& Schild 1997) gives values of $H_0<70$
(Dahle, Maddox, \& Lilje 1994; Turner 1996); the most recent analysis
gives $H_0=63\pm 12$ at the 95\% confidence level (Kundi\'c et al.~1996). 
Further progress is expected
from narrow double quasars like B0218+357 (Patniak, Porcas \& Browne 1995)
because the deflecting mass model is here better constraint. Present data
give for the least mentioned object $\tau=12\pm 3$ days (Corbett et al.~1995)
and $H_0\approx 60$ (Nair 1995).
 
The measurement  of the \lq\lq Doppler peaks\rq\rq ~in the CMB fluctuation
spectrum opens a new dimension to the meaning of $H_0$. A comparison
of the inferred and observed values of $H_0$ will have a profound effect
on our understanding  of the very early universe. The first results, giving
$30<H_0<50$ (or at most 70), are most encouraging (Lasenby 1996).

\section{Conclusions}

It is now possible to determine $H_0$ along three lines of attack as
described in Sections 2, 4, and 5. The results are repeated in
Table~\ref{tab:results}. Also shown is the high-weight determination of
$H_0$ from SNe$\,$Ia (Saha 1996).

The agreement of the different determinations is so good that the
{\sl external} error of $H_0=55\pm 7 \,(\pm 13\%)$ seems to be secure
unless one postulates a conspiracy of all data.

It should be noted, however, that $H_0$ from field galaxies and SNe$\,$Ia
depends entirely on the calibration via Cepheids and that Cepheids carry
also much of the weight of the multiple distance determinations of the 
Virgo cluster leading directly to a large-scale value of $H_0$.

\begin{table}
\noindent \caption{Different determinations of $H_0$}
\label{tab:results}
\begin{center} \footnotesize
\begin{tabular}{lrcc}
%\noalign{\smallskip}
\hline
%\noalign{\smallskip}
method & range{\phantom{M}} & $H_0$ & external error \\
       & km s$^{-1}$ & km s$^{-1}$ Mpc$^{-1}$ & \\
%\noalign{\smallskip}
\hline
%\noalign{\smallskip}
field galaxies & 4000 & $53\pm 3$ & $\pm10$ \\
Virgo cluster & 10000 & $54\pm 4$ & $\pm 8$ \\
SNe$\,$Ia & 30000 & $58\pm 4$ & $\pm 6$ \\
physical methods & large & 50-65 & ($\pm 15$) \\
\hline
adopted & & 55 & $\pm 7$ \\
\end{tabular}
\end{center}
\end{table}

Yet Cepheids are the best and least controversial distance indicators.
The zeropoint of their period-luminosity (P-L) relation has moved by
only 0.07 mag over an interval of 30~years (Kraft 1961; Sandage \&
Tammann 1968; Feast \& Walker 1987; Madore \& Freedman 1991). The
calibration, using only Galactic Cepheids, gives a distance modulus of
LMC of $(m-M)=18.50\pm 0.10$ (Sandage \& Tammann 1968; Feast \& Walker
1987), which is independently confirmed to better than 0.10 mag
through the shell of the SN 1987A in LMC (Panagia et al.~1996),
RR~Lyr stars [(Walker (1993) with the calibration of Sandage (1993c)],
the red-giant tip (Lee, Freedman, \& Madore 1993), the Baade-Becker-Wesselink 
method in $BVIJHK$ for RR~Lyr stars (Laney \& Stobie 1992), and the $I$ 
magnitude diameters of Cepheids (di Benedetto 1995).

The slope of the P-L relation taken from LMC is uncritical as long as
the available Cepheids cover a sufficient period interval. Metallicity
effects on the P-L relation are small (Freedman \& Madore 1990; Chiosi, Wood,
\& Capitanio 1993; Sandage 1996b); any such effects enter with low weight
because no strongly metal-deficient galaxies are considered here.
Selection bias (Sandage 1988b) can be avoided if the
Cepheids span a sufficient period interval.

These or similar arguments have led to a general consent that
Cepheids, as far as they can reliably be observed, are the most
fundamental distance indicators known. The zero point error of the
extragalactic distance scale introduced by the Cepheid P-L relation
is hence certainly {\sl less} than 10\% (0.2 mag). The concordant 
evidence from stepping up the distance scale through the Virgo cluster,
field galaxies, and SN$\,$Ia to the large scale value of $H_0$ makes
it unlikely that the final error could be larger than 13\%. 
 
\begin{acknowledgments}
{\bf Acknowledgments.} The authors thank Dr.~R.~Giovanelli for
pre-publication data.  They are indebted to Dres.~D.~Branch and
A.~Sandage for many stimulating discussions and they thank the
organizers for this excellent meeting. They gratefully acknowledge
support of the Swiss National Science Foundation.
\end{acknowledgments}

\end{document}